\documentclass[a4paper, oneside, twocolumn, notitlepage, 10pt]{extarticle_ecoc}
\usepackage{ecoc}
\usepackage{comment}
\usepackage{color}
\usepackage{subfigure}
\usepackage{enumerate}
\usepackage{enumitem}
\usepackage{stfloats}
\usepackage{psfrag}
\usepackage{setspace}
\usepackage{balance}
\usepackage{enumitem}
\usepackage{tikz}
\usepackage{pgfplots,amsfonts}
\usepackage{pgfplotstable}
\usepackage{amsmath,amssymb}
\usepackage{xcolor}
\usepackage{graphicx}
\usepackage{tikz}
\usepackage{pgfplots}
\usepackage{float}
\usepackage{balance}
\usepackage{setspace}
\usepackage{xspace}
\usepackage[normalem]{ulem}

\usepackage{dsfont,empheq}
\usepackage{multicol,multirow}

\usepackage{booktabs, cellspace, hhline}

\usetikzlibrary{shapes}
\usetikzlibrary{spy}
\usetikzlibrary{fit}
\usetikzlibrary{shapes.multipart}
\usetikzlibrary{positioning}
\pgfplotsset{compat=1.14}
\usepackage{wrapfig}
\usepackage[acronym,nomain]{glossaries}
\usepackage{arydshln}
\usepackage{booktabs,makecell,multirow,tabularx}
\usetikzlibrary{calc}

\begin{document}
\selectlanguage{english}    


\title{Experimental Demonstration of Rate-Adaptation via Hybrid Polar-BCH Product Code for Flexible PON}%


\author{
    Yifan~Ye\textsuperscript{(1)}, Bin~Chen\textsuperscript{(1),}*,
    Xiang~Li\textsuperscript{(2)}, Yi~Lei\textsuperscript{(1)}, Zhiwei~Liang\textsuperscript{(1)}, Qingqing~Hu\textsuperscript{(1)}, Can~Zhao\textsuperscript{(1)}, Yanni~Ou\textsuperscript{(3)}
}

\maketitle                  


\begin{strip}
    \begin{author_descr}

        \textsuperscript{(1)} School of Computer Science and Information Engineering, Hefei University of Technology, Hefei, China
        *  \textcolor{blue}{\uline{bin.chen@hfut.edu.cn}}

        \textsuperscript{(2)} School of Mechanical Engineering and Electronic Information, China University of Geosciences, Wuhan, China
        
        \textsuperscript{(3)}  State Key Laboratory of Information Photonics and Optical Communications, Beijing University of Posts and Telecommunications, Beijing, China

    \end{author_descr}
\end{strip}

\renewcommand\footnotemark{}
\renewcommand\footnoterule{}


\begin{strip}
    \begin{ecoc_abstract}
        The flexible-rate Polar-BCH product codes are experimentally demonstrated in a  coherent passive optical network system with 16QAM  for the first time. Using a new hybrid soft- and hard-decision decoder, we achieve a  power gain of  upto 1.75~dB over traditional BCH-BCH product codes after 48~km transmission. \textcopyright2025 The Author(s)
    \end{ecoc_abstract}
\end{strip}


\section{Introduction}
The surge of data-heavy applications like cloud computing, ultra-HD streaming, and remote work is driving an unprecedented demand for flexible, high-capacity access networks. To meet these emerging demands, passive optical networks (PONs) have been evolving toward higher capacities and more flexible resource allocation strategies. 
To further enhance these capabilities, coherent detection has been introduced into PON systems, enabling advanced modulation formats and significantly improving receiver sensitivity \cite{8831407,10926245,10926203}. Building on this, recent research has explored flexible PON architectures, focusing on solutions such as hybrid transceivers and local oscillator (LO) power tuning \cite{10896776,9748848}, which enable adaptive resource management based on varying link conditions. Moreover, techniques like multi-format modulation, probabilistic shaping, and discrete multi-tone (DMT) \cite{9606115,9508838,9748265} further provide flexibility by allowing the system to adjust transmission rates to meet dynamic traffic demands. However, these flexible transmission strategies also increase the susceptibility to transmission errors under varying channel conditions. In this context, forward error correction (FEC) plays a pivotal role, enabling high-capacity and flexible PON systems to fully realize their potential by ensuring robustness against errors and enhancing overall system reliability.

Low-density parity-check (LDPC) codes, as a class of near-capacity error-correcting codes with flexible rate compatibility, have been widely used in PON systems \cite{9768872,8946557}. However, when the code rate of LDPC codes deviates significantly from its original value due to puncturing or shortening, the gap between error performance and channel capacity widens, leading to reduced coding gain and poor adaptability in dynamic rate scenarios \cite{8840015}. In addition, the high decoding complexity of LDPC codes poses further challenges for real-time applications. Motivated by these challenges, product-like codes with various component codes and decoding methods have attracted attention due to their low complexity and strong error correction capability \cite{1089859,6074908,9685645,8856224,janz2025softoutputcoveredspacedecoding}. To provide more flexible rate adaptation without sacrificing performance, a novel coding scheme combining systematic polar and BCH codes with an efficient hybrid decoding algorithm was proposed recently \cite{yw}, offering a promising solution for achieving high capacity and flexible rates.



In this paper, we demonstrate the coding gains and flexibility  of a Polar-BCH product code (PC) with a hybrid soft- and hard-decision decoding algorithm   via  a 200G coherent  PON experiment validation. The experimental results demonstrate that the proposed scheme achieves a received power gain of up to 1.75~dB and 1.40~dB over BCH-BCH product code  with iterative bounded distance decoder (iBDD) \cite{1089859} and soft-aided bit-marking (SABM) decoder \cite{8856224}, respectively.
\vspace{-0.5em}
\section{Proposed Hybrid Product Code and Decoder}
Conventional product codes are constructed by arranging information bits into a two-dimensional matrix, with identical component codes applied along both rows and columns, forming a symmetric structure. The overall code rate is given by the product of the individual component code rates. Decoding is typically performed iteratively through independent hard-decision decoding on rows and columns, alternating between the two directions to progressively correct errors. Due to the reliance on hard decisions, the decoding performance is largely constrained by the capabilities of the component codes themselves. To enhance performance, some schemes introduce soft information to assist the hard-decision decoding process \cite{9685645,8856224}. However, the improvement is limited by the component codes' intrinsic capabilities and the nature of the decoding algorithm.
%

\begin{figure*}[!tb]
    \centering
    \includegraphics[width=1\textwidth]{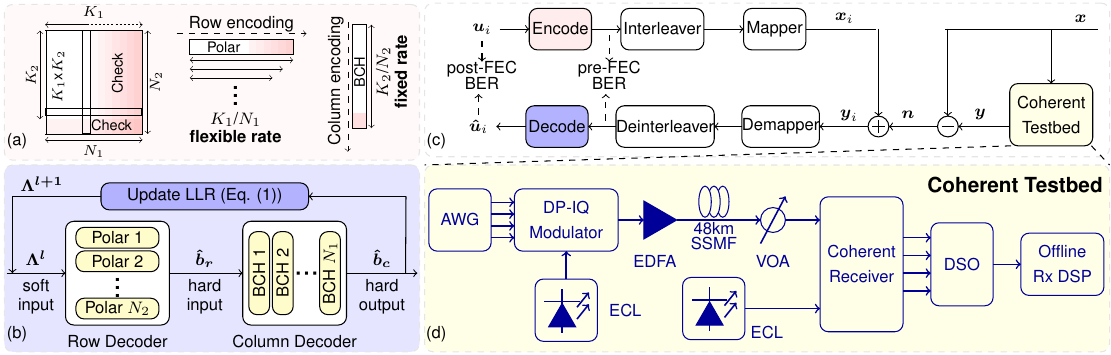}
    \caption{(a) Hybrid Polar-BCH product code structure.(b) Hybrid soft- and hard-decision decoding algorithm.(c) Block diagram of the FEC BER evaluation.(d) Experimental setup for the coherent PON based on dual-polarization 25 GBaud 16QAM.}
    \label{setup}
\end{figure*}

Motivated by the performance limitations of conventional symmetric product codes, we propose a hybrid product code structure that balances the performance of soft-decision decoding with the low complexity of hard-decision decoding \cite{yw}. 
The encoding structure is shown in Fig.~\ref{setup}(a). 
In this hybrid product code structure, Polar codes are employed as row component codes, while BCH codes serve as column component codes. The information bits are organized as a $K_1 \times K_2$ matrix and first encoded row-wise with Polar codes, followed by column-wise BCH encoding. Thanks to the inherent flexibility of Polar codes in rate configuration, various transmission requirements can be accommodated by adjusting the number of information bits $K_1$ under a fixed codeword length $N_1$, yielding a flexible rate of $K_1/N_1$.
By contrast, BCH codes operate at a fixed code rate of $K_2/N_2$, resulting in an overall code rate of $(K_1 \times K_2)/ (N_1 \times N_2)$. 
This asymmetric construction takes advantage of the complementary strengths of the two component codes: Polar codes exhibit excellent soft-input decoding performance under successive cancellation list (SCL) decoding \cite{7055304}, while BCH codes support efficient and reliable hard-decision decoding through BDD. 
The resulting $N_1 \times N_2$ code block is then modulated and transmitted over the channel.

The proposed hybrid soft- and hard-decision decoding (HSHD) algorithm  for the asymmetric PC structure is  illustrated in Fig.~\ref{setup}(b). 
The soft information $\boldsymbol{\Lambda}$ from the channel is first passed to the row decoder, where parallel SCL decoding is executed. The decoder
 outputs are first converted into hard decision bits and then  passed to the column BCH decoder, which performs BDD to generate the hard outputs $\boldsymbol{\hat b_c}$. 
 Since BCH decoding can reliably correct errors within its bounded distance, its outputs are generally considered more trustworthy, especially when Polar decoding yields uncertain soft information. 
 Therefore, when a conflict occurs between row and column decoding results, the soft information at the conflicting positions is updated based on the BCH decoding decision for the next polar decoding iteration.

During  the iterative  decoding,  the number of row and column decoding are implemented up to $l_{\max}$ iterations. 
After each iteration, the outputs $\boldsymbol{\hat b_r}$ and $\boldsymbol{\hat b_c}$ are compared. 
If they are identical, decoding terminates; otherwise, the soft information at the conflicting positions $(i, j)$ is updated as
\begin{align}\label{decode} 
\boldsymbol{\Lambda}^{l+1}(i,j)=\boldsymbol{\Lambda}^{l}(i,j)+\alpha\cdot(-1)^{\boldsymbol{\hat b_c}}, 
\end{align} 
where $\alpha$ is a scaling factor that controls the update magnitude. The updated soft information $\boldsymbol{\Lambda}^{l+1}$ is then used for the next iteration of Polar decoding. This design fully leverages the parallel decoding capability of Polar codes and the deterministic correction strength of BCH codes. We set $\alpha = 3$ to balance the influence of the BCH corrections and the soft information. By introducing the scaling factor $\alpha$, the scheme enables dynamic interaction between soft and hard decision information. Through this iterative process, the two sources of information reinforce each other: polar decoding benefits from BCH-based corrections, while BCH decoding gains more accurate hard inputs as iterations proceed. This cooperation  enables better performance at lower received power, enhancing error correction performance while maintaining low decoding complexity.
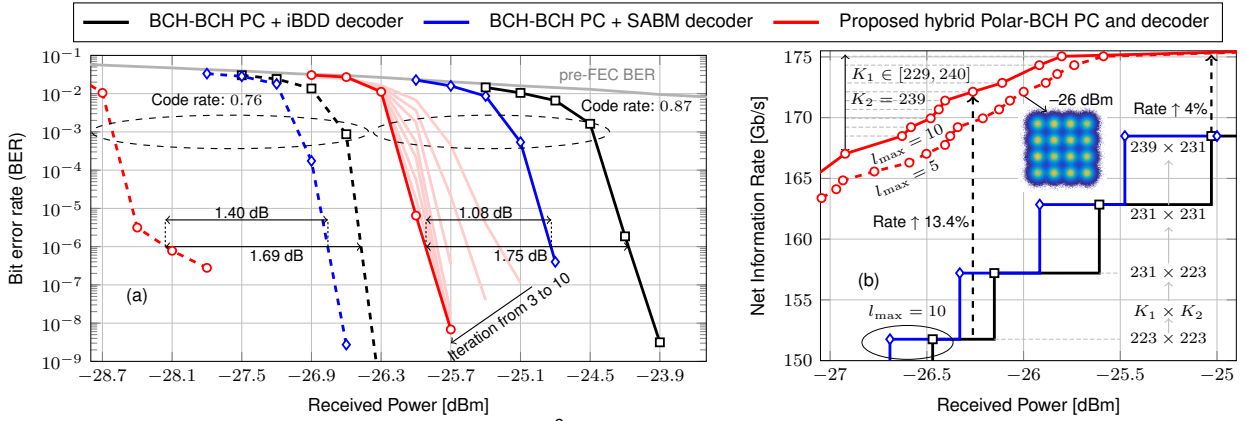
\begin{figure*}[!tb]
    \centering
    \begin{tikzpicture}
    \node at (0,0) {
        \begin{tikzpicture}
        \begin{axis}[
            hide axis,
            xmin=0, xmax=1,
            ymin=0, ymax=1,
            legend columns=3,
            legend style={
                at={(0.5,1.0)},
                anchor=south,
                font=\scriptsize,
                column sep=0.5em
            }
        ]
        \addlegendimage{black,line width=1.25pt}
        \addlegendimage{blue,line width=1.25pt}
        \addlegendimage{red,line width=1.25pt}
        \legend{
            BCH-BCH PC + iBDD decoder,
            BCH-BCH PC + SABM decoder,
            Proposed hybrid Polar-BCH PC and decoder
        }
        \end{axis}
        \end{tikzpicture}
    };
    \end{tikzpicture}

    \vspace{-1em} 
    \makeatletter
\renewcommand{\@thesubfigure}{\hskip\subfiglabelskip}
\makeatother
    \subfigure[]{
        \hspace{-2em} 
       \includegraphics[width=0.58\textwidth]{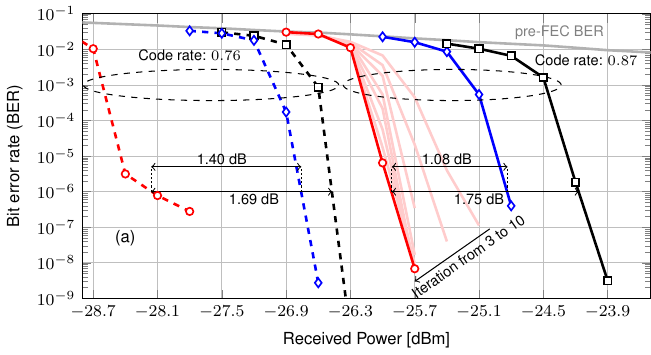}
    }
    \hspace{0.3em} 
    \subfigure[]{
        \includegraphics[width=0.41\textwidth]{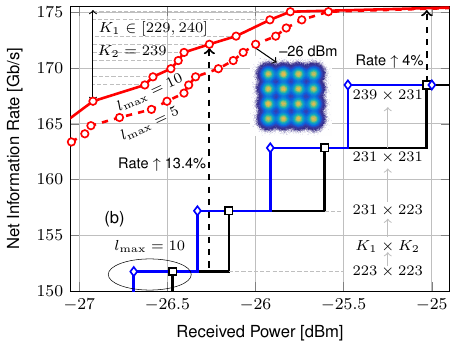}
    }
    \vspace{-2em} 
    \caption{(a) BER vs. received power of the (256, 239)\textsuperscript{2} codes with $l_{\max} = 10$, comparing two code structures (Polar-BCH and BCH-BCH) with three decoding schemes; (b) Net information rates at a target BER of $10^{-6}$ for considered PCs schemes.}
    \label{fig:combined}
\end{figure*}

\vspace{-0.5em}
\section{Experimental Setup and Results}
To evaluate the proposed hybrid coding scheme,  the system  we used is illustrated in Fig.~\ref{setup}(c). 
Random symbol sequences $\boldsymbol{x}$ are generated and transmitted through a coherent testbed (see Fig.~\ref{setup}(d)) to obtain the received signal $\boldsymbol{y}$.
The noise vector $\boldsymbol{n}$ is generated from  the received signal $\boldsymbol{y}$ and the transmitted signal $\boldsymbol{x}$, representing the total disturbances observed during transmission (such as quantization errors, photonic noise, amplified spontaneous emission (ASE) noise, etc.), and can be regarded as a sample vector of channel noise.

The original information bit sequence $\boldsymbol{u}_i$ is encoded and interleaved, then mapped to modulation symbols $\boldsymbol{x}_i$, which are added with the noise sample vector $\boldsymbol{n}$ to obtain the received signal $\boldsymbol{y}_i$. Demodulation, deinterleaving, and decoding are then performed to recover the estimated bit sequence $\boldsymbol{\hat{u}}$, which is compared with $\boldsymbol{u}_i$ to evaluate the post-FEC BER performance.

The experiment coherent testbed is shown as Fig.~\ref{setup}(d). A 1550 nm external cavity laser (ECL) with a linewidth of 100~kHz is used to modulate waveforms generated by a four-channel arbitrary waveform generator (AWG). 
The 25 GBaud 16QAM signal was then pre-equalized and resampled at 120 GSa to align with the AWG's  sampling rate. 
Then the electrical signal was converted into an optical signal using a dual-polarization IQ modulator and   amplified by an erbium-doped fiber amplifier (EDFA) before transmitting over a 48~km standard single-mode fiber (SSMF). A variable optical attenuator (VOA) was employed to adjust the received optical power level. 
At the receiver, the optical signal was captured by an integrated coherent receiver, using the ECL as a local oscillator (LO). The four outputs from the coherent receiver were sampled with an 80 GSa/s real-time digital storage oscilloscope (DSO).

To evaluate the post-FEC BER performance of a coherent PON system, the iBDD- and SABM-decoded BCH-BCH PC and  the proposed Polar-BCH PC with hybrid FEC are evaluated with the maximum iterations of $l_{\text{max}}=10$, as shown in Fig.~\ref{fig:combined}.
In Fig.~\ref{fig:combined}(a), we can observe that the proposed Polar-BCH PC achieves gains of 1.75~dB and 1.08~dB over iBDD and SABM, respectively, at a BER of $10^{-6}$ with a code rate of 0.87, and gains of 1.69~dB and 1.40~dB with a code rate of 0.76.
These results demonstrate the superior  error correction capability of the proposed scheme.
The coding  gain stems not only from the inherent advantages of soft-decodable component codes like Polar, but also from the strategic integration of BCH decoding outcomes into the iterative process. Specifically, soft decisions are incorporated along the horizontal decoding direction, improving the accuracy of information exchange between decoders. Moreover, when conflicts occur, the more reliable BCH-based column decoding results are used to update the soft information of the row decoder, effectively guiding the iterative process and mitigating error propagation.



Fig.~\ref{fig:combined}(b) illustrates the net information rate 
under different received power for BCH-BCH PC with two decoding schemes and Polar-BCH PC with proposed hybrid decoder, at a post-FEC BER of $10^{-6}$. 
As the received power increases, the system can support higher code rates and switches to the next higher rate level once the target BER is just satisfied.

%

Compared to BCH-BCH PCs with varying $K_1 \times K_2$ information bits, where the adjustment process appears as a stair-step progression due to the discrete nature of hard codes, the Polar-BCH product code enables smoother adaptation by incorporating a Polar code with flexible information length as the row component. This trend is clearly observed in Fig.~\ref{fig:combined}(b), where increasing the Polar code information length $K_1$ from 229 to 240 (with the BCH component $K_2$ fixed at 239) results in a net information rate improvement of 4\% to 13.4\% under the same received power conditions. Such performance gain highlights the advantage of finer rate granularity enabled by the hybrid design. The resulting flexibility allows for more continuous rate adjustment, better matching channel variations while maintaining the target BER. Moreover, such adaptability is particularly valuable in dynamic access networks, enabling flexible resource allocation by adjusting the ONU's transmission rate based on user location, traffic demand, or link quality, thus enhancing overall efficiency and quality of service.

\vspace{-0.5em}
\section{Conclusions}
We experimentally validated the  performance of the Polar-BCH product code with hybrid FEC in a flexible-rate dual-polarization 16QAM transmission over 48~km, achieving a power gain of up to 1.75~dB over traditional BCH-BCH product codes. 
This approach can support flexible code rate adaptation without requiring complex modification to the existing PON architecture.



\section{Acknowledgements}
This work is supported by the NSFC Program (No.~62171175), the Fundamental Research Funds for the Central Universities under Grant JZ2024HGTG0312.

\printbibliography


\end{document}